# Average and instantaneous velocities of energy of evanescent modes

**Zhi-Yong Wang[*], Wen-Chao Wang, Qi Qiu, Cai-Dong Xiong, Liu Yong**

*School of Optoelectronic Information, University of Electronic Science and Technology of China,*

*Chengdu 610054, CHINA*

Many theoretical and experimental investigations have presented a conclusion that evanescent electromagnetic modes can superluminally propagate. However, in this paper, we show that the average energy velocity of evanescent modes inside a cut-off waveguide is always less than or equal to the velocity of light in vacuum, while the instantaneous energy velocity can be superluminal, which does not violate causality according to quantum field theory: the fact that a particle can propagate over a space-like interval does preserve causality provided that here a measurement performed at one point cannot affect another measurement at a point separated from the first with a space-like interval.

## I. INTRODUCTION

How long does it take a particle to tunnel through a potential barrier? This is the issue of tunneling times that has been controversial for decades. Over the last 20 years, the investigations on the issue of tunneling times have experienced a strong stimulus by the results of analogous experiments with evanescent electromagnetic waves (also called *evanescent modes*) [1-6]. As a result, nowadays, there have been many theoretical and experimental reports on the propagation of evanescent electromagnetic modes at speeds exceeding the speed of light in vacuum [7-23], and the experimental results have shown that the phase time do describe the barrier traversal time [7].

However, to understand the physical meanings of these superluminal behaviors, there has given rise to much controversy. Part of the controversy stems from the fact that these superluminal velocities are the group velocities of evanescent modes. In view of which, in this paper we will study both an average and an instantaneous energy velocity of evanescent modes inside a cut-off waveguide, and show that the average velocity is always less than or equal to the velocity of light in vacuum, while the instantaneous one can be superluminal. Moreover, we show such results are in agreement with quantum field theory.

## II. AVERAGE ENERGY VELOCITY OF EVANESCENT MODES INSIDE A CUT-OFF WAVEGUIDE

As we know, a single purely evanescent mode cannot transmit any time averaged power, which is due to the fact that the electric and magnetic fields are in quadrature (90° out of phase). This is the case in an infinitely long evanescent region. In a finite region, however, there will be some reflection of the forward attenuating evanescent mode at the exit [24]. Because the cutoff waveguide has a purely imaginary characteristic impedance, the reflection coefficient will have a phase shift associated with it. The reflected backward evanescent mode has a phase shift so that the total electric field is no longer in quadrature with the magnetic field. Thus the interference between the forward and backward evanescent modes gives rise to a nonzero time average power flow. This power flow is not due to a propagating wave but can be seen as the beating between two evanescent cavity modes. In addition to the purely reactive pulsations of energy, there is a time averaged contribution due to the fact that net energy escapes through the boundary during each cycle.

Assume that a hollow rectangular waveguide is made of perfect conductors, choosing a Cartesian coordinate system ($x$, $y$, $z$) by taking the z-axis along the length of the waveguide, and the x- and y-axes parallel to the broad and narrow sides of the waveguide, respectively. The waveguide is a straight pipe with a part of reduced cross section from $z=0$ to $z=L$, its cross dimensions are $d \times h$ for $z<0$ and $z>L$, while $a \times b$ for $0 \leq z \leq L$, where $a>b$, $d>h$. Assume that one has $0 \leq x \leq a$ and $0 \leq y \leq b$ in our coordinate system. Let $d \rightarrow +\infty$, $h \rightarrow +\infty$, and then for *z*<0 and *z*>*L*, the cut-off frequency of the waveguide is taken to be zero, approximatively.

As we know, when a pulse of electromagnetic energy encounters a discontinuity in a waveguide, it is partially reflected and partially transmitted. In the process some of the energy may end up in higher-order evanescent modes which, being nonpropagating, are confined to the vicinity of the obstacle. Far away from the obstacle, its effects can be completely described by its reflection coefficient, its impedance, or its scattering matrix. A key simplification is that the scatterer does not alter the polarization state of the incident wave so that one may use scalar analysis. For simplicity, let us assume that the waveguide supports only the dominant $TE_{10}$ mode, and then the cut-off frequencies of the waveguide can be written as

$$\omega_c = \begin{cases} c\pi/d \approx 0, \text{ for } z<0 \text{ or } z>L \\ c\pi/a, \text{ for } 0<z<L \end{cases}, \quad (1)$$

where *c* is the velocity of light in vacuum. In our coordinate system, the electric and magnetic fields of the $TE_{10}$ mode can be written as $\boldsymbol{E}=(0,E_y,0)$ and $\boldsymbol{H}=(H_x,0,H_z)$, respectively. Let $\Psi=E_y,H_x,H_z$, these field quantities satisfy the wave equation

$$(\frac{d^2}{dz^2}+k^2-k_c^2)\Psi=0, \quad (2)$$

where $k_c=\omega_c/c$, $k=|\boldsymbol{k}|=\omega/c$, $\boldsymbol{k}$ is the wave number vector of the $TE_{10}$ mode, and $\omega$ is the frequency. In terms of $\beta \equiv -ik_z = \sqrt{k_c^2-k^2}$, the wave number vector can be expressed as $\boldsymbol{k}=(\pm k_c,0,i\beta)$. Assume that the frequency $\omega$ satisfies $c\pi/a > \omega > c\pi/d \approx 0$, and then the waveguide within the region of 0≤*z*≤*L* becomes a cut-off waveguide. From now on, let $\omega_c$ just denote the cut-off frequency of the cut-off waveguide (i.e., the waveguide within the region of 0≤*z*≤*L*), which implies that $\omega_c \equiv c\pi/a$ and $\omega_c > \omega > 0$. At the entrance of the cut-off waveguide (i.e., at $z=0$), an incident electromagnetic pulse (as the $TE_{10}$ mode) will be partially reflected and partially transmitted. Let $R$ and $T$ denote the reflection and transmission coefficients, respectively, they satisfy the normalization condition of $|R|^2+|T|^2=1$. For $z<0$ and $z>L$, the solutions of Eq. (2) can be rewritten as, formally

$$\Psi(x,z,t)=\begin{cases}\Phi(x,z,t)+R\Phi(x,z,t), & z<0 \\ T\Phi(x,z,t), & z>L\end{cases}. \qquad (3)$$

For the moment the cut-off waveguide plays the role of a photonic potential barrier. In view of the fact that, for the $TE_{10}$ mode, the waveguide can be imagined as just consisting of two vertical plates with the separation $d$ for $z<0$ and $z>L$, and with the separation $a$ for $0 \leq z \leq L$. Therefore, our discussions can be shown in Fig.1 (it depicts the waveguide's longitudinal profile that is parallel to the broadside of the waveguide).

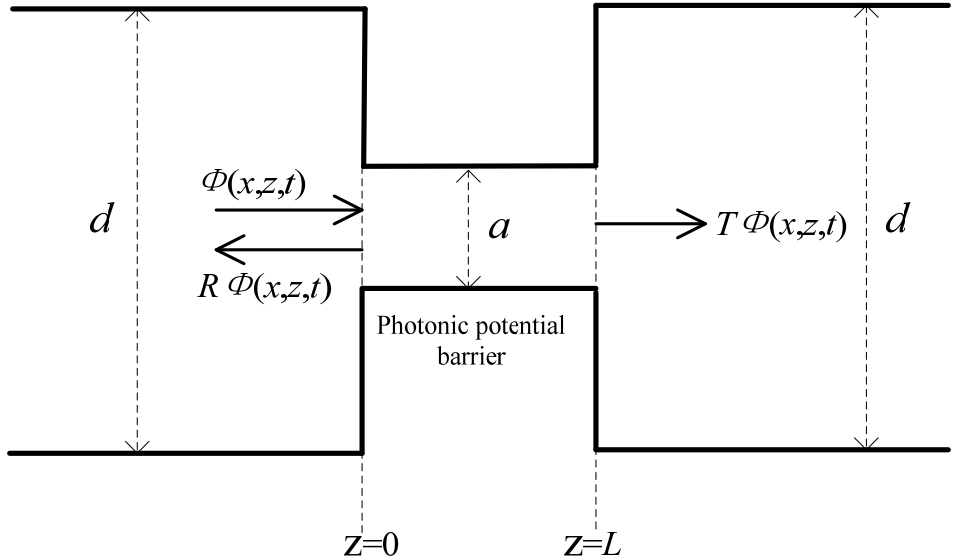

FIG. 1. Photonic potential barrier formed by the cut-off waveguide between $z=0$ and $z=L$. The frequency of the $TE_{10}$ mode satisfies $c\pi/a > \omega > c\pi/d \approx 0$ $(d\rightarrow+\infty)$.

At the entrance and exit of the cut-off waveguide (i.e., at $z=0$ and $z=L$), both the electromagnetic fields and their first derivatives satisfy the condition of continuity, such that for $0<z<L$ (i.e., inside the cut-off waveguide), one can prove that the electromagnetic fields, as the *evanescent* $TE_{10}$ modes, can be written as, respectively

$$\begin{cases} E_y = (iH_0\omega\mu_0/k_c)\sin k_c x \exp(-i\omega t)[A\exp(\beta z)+B\exp(-\beta z)] \\ H_x = (H_0\beta/k_c)\sin k_c x \exp(-i\omega t)[-A\exp(\beta z)+B\exp(-\beta z)] \\ H_z = H_0\cos k_c x \exp(-i\omega t)[A\exp(\beta z)+B\exp(-\beta z)] \end{cases} \quad (4)$$

where $\varepsilon_0$ and $\mu_0$ are respectively the vacuum permittivity and permeability satisfying $\mu_0\varepsilon_0 = 1/c^2$, $H_0$ is a constant amplitude, $\beta = \sqrt{k_c^2 - k^2}$, and

$$\begin{cases} T = \dfrac{-2i(k/\beta)\exp(-ikL)}{[1-(k/\beta)^2]\sinh\beta L - 2i(k/\beta)\cosh\beta L} \\ A = \dfrac{T}{2}[1+\dfrac{ik}{\beta}]\exp(ikL-\beta L) \\ B = \dfrac{T}{2}[1-\dfrac{ik}{\beta}]\exp(ikL+\beta L) \end{cases} \quad (5)$$

For the moment, inside the cut-off waveguide the Poynting vector $S_z$ and energy density $w$ are respectively,

$$S_z = \frac{1}{2}\text{Re}(\boldsymbol{E}\times\boldsymbol{H}^*)_z = \frac{1}{2}\text{Re}(-E_y H_x^*), \quad w = \frac{1}{4}(\varepsilon_0|E_y|^2 + \mu_0|H_x|^2 + \mu_0|H_z|^2). \quad (6)$$

The average velocity of energy inside the cut-off waveguide is

$$\bar{v}_e(z) = \int_0^a\int_0^b S_z \,dxdy \bigg/ \int_0^a\int_0^b w \,dxdy, \quad 0 < z < L, \quad (7)$$

Substituting Eqs. (4)-(6) into Eq. (7), one has

$$\begin{cases} \bar{v}_e(z) = \dfrac{P_z}{W} = c\dfrac{2k^2}{k_c^2}[\dfrac{k_c^2}{\beta^2}\cosh 2\beta(z-L) + \dfrac{k^2(\beta^2-k^2)}{k_c^2\beta^2}]^{-1} \\ P_z = \int_0^a\int_0^b S_z \,dxdy = |TH_0|^2 ab\mu_0 ck^2/4k_c^2 \\ W = \int_0^a\int_0^b w \,dxdy = \dfrac{1}{8}|TH_0|^2 ab\mu_0[\dfrac{k_c^2}{\beta^2}\cosh 2\beta(z-L) + \dfrac{k^2(\beta^2-k^2)}{k_c^2\beta^2}] \end{cases} \quad (8)$$

One can show that $|\bar{v}_e(z)| \leq c$, i.e., the average energy velocity of evanescent modes cannot be superluminal.

Nevertheless, some people might argue that, because the Poynting vector $S_z$ stands for the flow of momentum along the waveguide, to obtain the average energy velocity, one should calculate the energy density $w$ via the field quantities that make a contribution to the

Poynting vector $S_z$, that is

$$w = (\varepsilon_0 |E_y|^2 + \mu_0 |H_x|^2)/4 , \qquad (9)$$

which implies that the resulting average energy velocity is larger than the one given by Eq. (8), in particular, it might become superluminal. However, if Eq. (9) were right, using Eq. (9) one has

$$\begin{cases} \bar{v}_e(z) = \dfrac{P_z}{W} = c \dfrac{4k^2}{k_c^2} [\dfrac{k_c^2}{\beta^2} \cosh 2\beta(z-L) + \dfrac{(\beta^2 - k^2)^2}{k_c^2 \beta^2}]^{-1} \\ P_z = \int_0^a \int_0^b S_z \mathrm{d}x \mathrm{d}y = |TH_0|^2 ab\mu_0 ck^2 / 4k_c^2 \\ W = \int_0^a \int_0^b w \mathrm{d}x \mathrm{d}y = \dfrac{1}{16} |TH_0|^2 ab\mu_0 [\dfrac{k_c^2}{\beta^2} \cosh 2\beta(z-L) + \dfrac{(\beta^2 - k^2)^2}{k_c^2 \beta^2}] \end{cases} , \qquad (10)$$

using Eq. (10) one can show that the result of $|\bar{v}_e(z)| \leq c$ is also valid.

Therefore, the average energy velocity of evanescent modes is always subluminal (or equal to c), which implies that for expectation values or ensemble average, Einstein causality is preserved.

It is worthy of note that the energy velocity of evanescent modes inside a cutoff waveguide has also been studied in Ref. [24]. However, some results in Ref. [24] might be questionable. For example, the equation (38) in Ref. [24] is not valid for the energy velocity of evanescent modes, which is due to the fact that the evanescent modes satisfy $\omega < \omega_1$ such that the energy velocity becomes an imaginary number.

## III. INSTANTANEOUS ENERGY VELOCITY OF EVANESCENT MODES INSIDE A CUT-OFF WAVEGUIDE

To investigate the instantaneous energy velocity of evanescent modes inside the above-mentioned cutoff waveguide, let us study the probability for a photon to propagate a space-like interval along the cut-off waveguide. In Chapter III, the natural units of measurement ($\hbar = c = 1$) is applied, repeated indices must be summed according to the Einstein rule, and the space-time metric tensor is chosen as $g^{\mu\nu} = \text{diag}(1,-1,-1,-1)$, $\mu,\nu = 0,1,2,3$, $x^{\mu} = (t,\boldsymbol{x}) = (t,x,y,z)$, and so on.

Let $|0\rangle$ denote the vacuum state of a field operator $\varphi(x^{\mu}) = \varphi(t,x,y,z)$, according to quantum field theory, the quantity $G(x^{\mu} - x'^{\mu}) \equiv \langle 0|\varphi(x^{\mu})\varphi(x'^{\mu})|0\rangle$ represents a transition probability amplitude from the quantum state of $\varphi(x'^{\mu})|0\rangle$ to the one of $\varphi(x^{\mu})|0\rangle$, such that $|G(x^{\mu} - x'^{\mu})|^2$ is associated with the probability for a particle to propagate over a spacetime interval of $(x^{\mu} - x'^{\mu})$. From now on, all the field quantities inside the waveguide become operators in the field-quantization level. In terms of the electric field operator $E_y$ inside the cut-off waveguide, let us study the quantity $G(x^{\mu} - x'^{\mu}) \equiv \langle 0|E_y(x^{\mu})E_y(x'^{\mu})|0\rangle$ without loss of generality. Likewise, the function $G(x^{\mu} - x'^{\mu})$ is related to the probability amplitude for photons to propagate from $x'^{\mu} = (t',x',y',z')$ to $x^{\mu} = (t,x,y,z)$ inside the waveguide, it can be regarded as a correlation function for photons inside the cut-off waveguide, too. If $G(x^{\mu} - x'^{\mu}) \neq 0$ for a space-like interval $(x^{\mu} - x'^{\mu})$, the probability for photons to propagate superluminally along the cut-off waveguide does not vanish, such that the instantaneous energy velocity of evanescent modes can be superluminal. In the present case, the 4-dimensional wave-number vector of photons in the free space inside the waveguide is $k^{\mu} = (\omega, k_x, k_y, k_z) = (\omega, \pi/a, 0, k_z)$, that is, $k_x = \omega_c = \pi/a$ (note that the natural units of measurement ($\hbar = c = 1$) is applied). Applying QED and combining with our conditions, one can obtain

$$G(x^{\mu} - x'^{\mu}) = (\frac{\partial^2}{\partial y^2} - \frac{\partial^2}{\partial t^2})D(x^{\mu} - x'^{\mu}), \qquad (11)$$

where

$$D(x^\mu - x'^\mu) = C\int \frac{dk_z}{2\pi} \frac{1}{2\omega} \exp[-i\omega(t-t') + ik_z(z-z')], \quad (12)$$

where $\omega = \sqrt{\omega_c^2 + k_z^2}$ is the frequency of photons and $C = \exp[i(\pi/a)(x-x')]/(2\pi)^2$ is a phase factor. Using Eqs. (11) and (12) one has

$$G(x^\mu - x'^\mu) = -\frac{\partial^2}{\partial t^2} D(x^\mu - x'^\mu), \quad (13)$$

For the cut-off waveguide, one has $0 < \omega = \sqrt{\omega_c^2 + k_z^2} < \omega_c$, which implies that $k_z = i\beta$ ($\beta$ is a real number satisfying $-\omega_c < \beta < \omega_c$). Seeing that the waveguide is placed along the z axis, let us take $x'^\mu = (0,0,0,0)$ and $x^\mu = (t,0,0,z)$ with $t, z \geq 0$ for simplicity, and then one has $C = 1/(2\pi)^2$. By substituting $x'^\mu = (0,0,0,0)$, $x^\mu = (t,0,0,z)$, $k_z = i\beta$, $0 < \omega < \omega_c$ and Eq. (12) into Eq. (13), one can obtain

$$G(x^\mu) = (1/16\pi^3)\int_0^{\omega_c} d\beta \sqrt{\omega_c^2 - \beta^2} \exp(-i\sqrt{\omega_c^2 - \beta^2}\, t - \beta z). \quad (14)$$

Applying the integral representation of the Hankel function of the second kind

$$H_0^{(2)}(r) \equiv \frac{2}{\pi} \int_0^{\pi/2} d\theta \exp(-ir\sin\theta), \quad (15)$$

one can obtain (notice that $x^\mu x_\mu = c^2 t^2 - z^2 = t^2 - z^2$ in the natural units of $\hbar = c = 1$):

$$G(x^\mu) = \frac{\omega_c}{32\pi^2 \sqrt{x^\mu x_\mu}}[H_1^{(2)}(\omega_c\sqrt{x^\mu x_\mu}) - tH_2^{(2)}(\omega_c\sqrt{x^\mu x_\mu})]. \quad (16)$$

Obviously, $G(x^\mu - x'^\mu) = G(x^\mu) \neq 0$ for the space-like interval of $x^\mu x_\mu < 0$. As we know, for a timelike interval $x^\mu x_\mu = t^2 - z^2 > 0$, there is always an inertial frame in which $z = 0$; while, for a spacelike interval $x^\mu x_\mu = t^2 - z^2 < 0$, there is always an inertial frame in which $t = 0$. Then one can show the asymptotic behaviors of $G(x^\mu)$ as follows:

$$G(x^\mu) \sim \begin{cases} (t)^{-1/2} \exp(-i\omega_c t), & \text{as timelike interval } x^\mu x_\mu = t^2 \to +\infty, \\ (z)^{-3/2} \exp(-\omega_c z), & \text{as spacelike interval } x^\mu x_\mu = -z^2 \to -\infty. \end{cases} \quad (17)$$

Therefore, the probability for photons to propagate superluminally along the cut-off waveguide does not vanish, which implies that the instantaneous energy velocity of the evanescent modes can be superluminal.

## IV. CONCLUSIONS AND DISCUSSIONS

Up to now, we have come to the conclusion that the instantaneous energy velocity of evanescent modes can be superluminal, while the average energy velocity is always subluminal (or equal to c). However, such superluminal behavior does not violate causality. In fact, according to quantum field theory (see for example, Refs. [25] and [26]), a non-zero propagator or non-zero transition probability amplitude for a spacelike interval implies that a particle can propagate over the spacelike interval, but this spacelike propagation does not destroy causality provided that the commutator of two observables with a spacelike interval vanishes. In other words, causality is maintained as long as a measurement performed at one point cannot affect a measurement at another point whose separation from the first is spacelike.

On the other hand, just as S. Weinberg discussed in Ref. [27], although the relativity of temporal order raises no problems for classical physics, it plays a profound role in quantum theories. The uncertainty principle tells us that when we specify that a particle is at position $x_1$ at time $t_1$, we cannot also define its velocity precisely. In consequence there is a certain chance of a particle getting from $(t_1, x_1)$ to $(t_2, x_2)$ even if the spacetime interval is spacelike, that is, $|x_1 - x_2| > c|t_1 - t_2|$. To be more precise, the probability of a particle reaching $(t_2, x_2)$ if it starts at $(t_1, x_1)$ is nonnegligible as long as

$$0 < (x_1 - x_2)^2 - c^2(t_1 - t_2)^2 \leq (\hbar/mc)^2, \tag{18}$$

where ℏ is Planck's constant (divided by 2π), and *m* the particle's mass (and then ℏ/mc is the particle's Compton wavelength). We are thus faced with our paradox: if one observer sees a particle emitted at $(t_1, x_1)$, and absorbed at $(t_2, x_2)$, and if $(x_1 - x_2)^2 - c^2(t_1 - t_2)^2$ is positive (but less than or equal to $(\hbar/mc)^2$), then a second observer may see the particle absorbed at $x_2$ at a time $t_2$ before the time $t_1$ it is emitted at $x_1$. There is only one known way out of this paradox. The second observer must see a particle emitted at $x_2$ and absorbed at $x_1$. But in general the particle seen by the second observer will then necessarily be different from that seen by the first observer (it is the antiparticle of the particle seen by the first observer). In other words, to avoid a possible causality paradox, one can resort to the particle-antiparticle symmetry. The process of a particle created at $(t_1, x_1)$ and annihilated at $(t_2, x_2)$ as observed in a frame of reference, is identical with that of an antiparticle created at $(t_2, x_2)$ and annihilated at $(t_1, x_1)$ as observed in another frame of reference.

In our case, the antiparticle of the photon is the photon itself. Therefore, the process that a photon propagates superluminally from A to B as observed in a frame of reference, is equivalent to the process that the photon propagates superluminally from B to A as observed in another frame of reference, where causality is preserved provided that the measuring results from the same observer are consistent and do not conflict with any physical law. More generally, general relativity tells us that causality is always preserved provided that there does not form a closed timelike curve. In fact, one can define an effective rest mass of guided photons inside a waveguide as $m = \hbar\omega_c/c^2$ ($\omega_c$ is the cut-off frequency of the waveguide), and then the Compton wavelength of guided photons is $\lambdabar_c = \hbar/mc$ [28]. The walls of a cut-off waveguide localizes photons along the cross direction with a greater precision than the Compton wavelength of the photons, such that the quantum-mechanical effects cannot be ignored. As a result, the superluminal behavior of evanescent modes attributes to a purely quantum-mechanical effect, and it preserves quantum-mechanical causality or the so-called weak causality [29], that is, it preserves Einstein causality for expectation values or ensemble average only, not for individual probabilistic process.

**ACKNOWLEDGMENTS**

This work was supported by the National Nature Science Foundation of China (No. 60925019, 61090393). The first author (Z. Y. Wang) would like to greatly thank Prof. G. Nimtz for his helpful discussions. Moreover, we would like to greatly thank the referee for his helpful comments and valuable suggestions.

______________________________________________________________

*E-mail:   zywang@uestc.edu.cn

[1] M. Büttiker and R. Landauer,   Phys. Rev. Lett. **49**, 1739 (1982).

[2] E. H. Hauge and J. A. Støvneng, Rev. Mod. Phys. **61**, 917 (1989).

[3] V. S. Olkhovsky and E. Recami, Phys. Reports **214**, 339 (1992).

[4] R. Landauer and Th. Martin, Rev. Mod. Phys. **66**, 217 (1994).

[5] P. G. Kwiat, R. Y. Chiao and A. M. Steinberg, Physica B **175**, 257 (1991).

[6] Th. Martin and R. Landauer, Phys. Rev. A **45**, 2611 (1992).

[7] A. Enders and G. Nimtz J. Phys. I (France) **2**, 1693 (1992).

[8] A. Enders and G. Nimtz Phys. Rev. E **48**, 632 (1993).

[9] A. M. Steinberg, P. G. Kwiat, and R. Y. Chiao, Phys. Rev. Lett.**71**, 708 (1993).

[10] Ch. Spielmann, R. Szipöcs, A. Stingl, and F. Krausz, Phys. Rev. Lett. **73**, 2308 (1994).

[11] J. J. Carey, J. Zawadzka, D. A. Jaroszynski, and K. Wynne, Phys. Rev. Lett. **84**, 1431 (2000).

[12] A. Haibel, G. Nimtz, and A. A. Stahlhofen, Phys. Rev. E **63**, 047601 (2001).

[13] G. Nimtz and W. Heitmann, Prog Quant Electr. **21**, 81 (1997).

[14] G. Nimtz, A. Haibel, and R.-M. Vetter, Phys. Rev. E **66**, 037602 (2002).

[15] A. P. Barbero, H. E. Hernández-Figueroa and E. Recami, Phys. Rev. E **62**, 8628 (2000).

[16] S. Longhi, P. Laporta, M. Belmonte, and E. Recami, Phys. Rev. E **65**, 046610 (2002).

[17] Zhi-Yong Wang, Cai-Dong Xiong, and B. He, Phys. Rev. A **75**, 013813 (2007).

[18] Zhi-Yong Wang and Cai-Dong Xiong, Phys. Rev. A **75**, 042105 (2007).

[19] D. Mugnai, A. Ranfagni and L. S. Schulman, Phys. Rev. E **55**, 3593 (1997).

[20] D. Sokolovski, Phys. Rev. A **81**, 042115 (2010).

[21] R. A. Sepkhanov, M. V. Medvedyeva, and C. W. J. Beenakker, Phys. Rev. B **80**, 245433 (2009).

[22] Ph. Balcou and L. Dutriaux, Phys. Rev. Lett.**78**, 851 (1997).

[23] I. Alexeev, K. Y. Kim and H. M. Milchberg, Phys. Rev. Lett. **88**, 073901 (2002).

[24] Herbert G. Winful, Phys. Rev. E **68**, 016615 (2003).

[25] M. E. Peskin and D. V. Schroeder, *An Introduction to Quantum Field Theory* (Addison-Wesley, New York, 1995), pp. 27-29.

[26] W. Greiner and J. Reinhardt, *Field Quantization* (Springer-Verlag, Berlin, 1996), p. 209.

[27] S. Weinberg, *Gravitation and Cosmology* (John Wiley & Sons, New York,1972), pp. 61-63.

[28] Wang Zhi-Yong, Xiong Cai-Dong and He Bing, Chin. Phys. B **17**, 3985 (2008).

[29] G. C. Hegerfeldt, Ann. Phys. (Leipzig) **7**, 716 (1998).